%% file: IEEE-conference-template-062824.tex
\documentclass[conference]{IEEEtran}
\IEEEoverridecommandlockouts
\usepackage{acro}
\usepackage{amsmath,amssymb,amsfonts}
\usepackage{algorithm}
\usepackage{algorithmic}
\usepackage{array}
\usepackage{balance}
\usepackage{booktabs}
\usepackage{caption}
\usepackage{cite}
\usepackage{comment}
\usepackage{graphicx}
\usepackage{hyperref}
\usepackage{listings}
\usepackage{siunitx}
\usepackage{subcaption}
\usepackage{tabularx}
\usepackage{textcomp}
\usepackage{xcolor}
\usepackage{tikz}
\usepackage{amsmath,amssymb,amsfonts}
\usepackage{algorithmic}
\usepackage{graphicx}
\usepackage{textcomp}
\usepackage{xcolor}
\usepackage{listings}
\usepackage{makecell}
\usepackage{pifont}
\usepackage{hyperref}
\usepackage{cite}
\usepackage{pgfplots}
\pgfplotsset{compat=1.18}
\usepackage{caption}
\usepackage{cleveref}

\DeclareAcronym{MBE}{
	short = MBE,
	long  = Model-based Engineering,
	sort  = abbrev,
}

\DeclareAcronym{LLM}{
	short = LLM,
	long  = Large Language Model,
	sort  = abbrev,
}

\DeclareAcronym{OMG}{
	short = OMG,
	long  = Object Management Group,
	sort  = abbrev,
}

\DeclareAcronym{MOF}{
	short = MOF,
	long  = Meta-Object Facility,
	sort  = abbrev,
}

\DeclareAcronym{EMF}{
	short = EMF,
	long  = Eclipse Modeling Framework,
	sort  = abbrev,
}

\begin{document}
	
    \title{Survey of GenAI for Automotive Software Development: From Requirements to Executable Code}
    

    \author{
    \IEEEauthorblockN{
    Nenad Petrovic, Vahid Zolfaghari, Andre Schamschurko, Sven Kirchner, Fengjunjie Pan, Chengdng Wu, \\
    Nils Purschke, Aleksei Velsh, Krzysztof Lebioda, Yinglei Song, Yi Zhang, Lukasz Mazur and Alois Knoll \vspace{0.2cm}
    }
    \IEEEauthorblockA{
    \textit{Chair of Robotics, Artificial Intelligence and Real-Time Systems} \\
    Technical University of Munich, Munich, Germany \\
    Email: \{nenad.petrovic, v.zolfaghari, andre.schamschurko, sven.kirchner, f.pan, chengdong.wu \\
    nils.purschke, aleksei.velsh, krzysztof.lebioda, yinglei.song, yi1228.zhang, lukasz.mazur, k\}@tum.de
    }
    }
    	
    \maketitle
    	
    \begin{abstract}
    Adoption of state-of-art Generative Artificial Intelligence (GenAI) aims to revolutionize many industrial areas by reducing the amount of human intervention needed and effort for handling complex underlying processes. Automotive software development is considered to be a significant area for GenAI adoption, taking into account lengthy and expensive procedures, resulting from the amount of requirements and strict standardization. In this paper, we explore the adoption of GenAI for various steps of automotive software development, mainly focusing on requirements handling, compliance aspects and code generation. Three GenAI-related technologies are covered within the state-of-art: Large Language Models (LLMs), Retrieval Augmented Generation (RAG), Vision Language Models (VLMs), as well as overview of adopted prompting techniques in case of code generation. Additionally, we also derive a generalized GenAI-aided automotive software development workflow based on our findings from this literature review. Finally, we include a summary of a survey outcome, which was conducted among our automotive industry partners regarding the type of GenAI tools used for their daily work activities.
    \end{abstract}

	\section{Introduction}
	\input{sections/01_introduction}

	\section{Background} \label{sec:background}
	\input{sections/02_background}

	\section{Methodology} \label{sec:methodology}
	\input{sections/02b_methodology}

	\section{Generalized workflow} \label{sec:workflow}
	\input{sections/03_workflow}

	\section{LLMs and RAG for requirements handling} \label{sec:rag}
	\input{sections/04_requirements}

	\section{LLMs and RAG for automotive regulation compliance} \label{sec:rag}
	\input{sections/04a_compliance}

	\section{LLM-based code generation} \label{sec:code}
	\input{sections/06_code_generation}

	\section{Tackling GenAI hallucinations} \label{sec:halucination}
	\input{sections/07_hallucination}

 	\section{Code analysis and optimization} \label{sec:optimization}
	\input{sections/08_code_optimization}

 	\section{VLM-based requirements summarization} \label{sec:vlm_req}
	\input{sections/09_vlm_reqs}

 	\section{Prompting techniques in automotive} \label{sec:prompt_tec}
	\input{sections/10_prompt}

 	\section{Survey: GenAI tools in automotive industry} \label{sec:survey}
	\input{sections/10_survey}

	\section{LLM usage overview} \label{sec:discussion}
	\input{sections/11_discussion}

	\section{Conclusion} \label{sec:conclusion}
	\input{sections/12_conclusion}
 
	\section{Acknowledgment} \label{sec:conclusion}
	\input{sections/13_ack}
	\bibliographystyle{IEEEtran}
	\balance
        \input{IEEE-conference-template-062824.bbl}

\end{document}

%% file: sections/01_introduction.tex
The breakthrough of Generative Artificial Intelligence (GenAI), especially Large Language Models (LLMs) in the last three years has a significant impact on many areas - from everyday routines to industry and manufacturing. LLMs exhibit strong capabilities when it comes to text summarization and generation and were found suitable for the automation of many human tasks. This affects many industrial domains \cite{Baptista2025}, among them the automotive industry \cite{staron2025} \cite{phatale2024} \cite{denninger2025}. Usually, the adoption of GenAI in the automotive and other industry domains target either to eliminate human intervention needed for repetitive tasks, speed-up complex processes and activities or even introduce newly added value by enabling novel use cases \cite{staron2025}.

The automotive industry is known for strict design, development, testing and manufacturing procedures which need to be compliant with numerous standards, making the time from research and development to production lines very long in practice. Therefore, innovation in this area is usually limited by time-consuming and expensive procedures which rely on domain-specific expertise and involve many manual steps. Additionally, the outcome must cover hundred thousands of requirements \cite{Habibullah2024}, even in the case of average-sized vehicles \cite{maier2023requirements}, making the development process even more complex. Based on the current works, it can be identified that GenAI adoption in automotive software development mostly targets requirements handling, compliance aspects, test scenario and code generation.

However, the adoption of GenAI solutions in practice brings many challenges in sensitive domains, \cite{matarazzo2025survey}, such as the automotive domain, with it. First, GenAI models are prone to so-called hallucinations - meaning they fantasize; often times in a believable way at first glance. This introduces an factor of uncertainty when it comes to the quality of the generated contents, which makes a direct unverified usage of the results almost impossible. For this reason, additional intermediate steps involving formally verified and grounded methods such as Model-Driven Engineering are used in combination.

Though, requirements in the automotive industry are highly valuable asset and their availability is often limited by legal means, such as non-disclosure agreements, making their processing more challenging. Therefore, it is usually considered that their exposing should be limited within organizational boundaries, making the usage of third party services and external cloud provider infrastructure limited. In this case, it is usually expected that smaller, locally deployable models are tailored, making them usable for narrow set of specific tasks, assuming that a sufficient amount of data was provided in the so-called fine-tuning process.

In this paper, we perform the survey of GenAI-based approaches applied for automation of crucial steps within automotive software development process, starting from requirements to code generation. Our goal is to identify underlying usage scenarios, underlying methods, models, as well as challenges for specific steps. The paper is structured as follows. In background section, we provide basics of the underlying GenAI-based methodology, covering Large Language Models (LLMs), prompting techniques targeting LLMs, Retrieval Augmented Generation (RAG) and Vision Language Models (VLMs). Additionally, we provide generalized workflow overview - from requirements to executable code, based on the existing literature, identifying the crucial steps where GenAI adoption is expected. After that, we consider each of this steps in details, providing the summary of the relevant works, as well as identifying the trends about the underlying GenAI models which are used. Finally, we also include the results of GenAI adoption survey performed in collaboration with industry partners from our projects.

%% file: sections/02_background.tex
\subsection{Large Language Models (LLMs)}
A Large Language Model (LLM) is a type of artificial intelligence model designed to understand and generate human language with remarkable proficiency. These models are trained on massive datasets of text, enabling them to grasp linguistic patterns, structure, and even subtle nuances. Typically, LLMs consist of billions of parameters. Modern LLMs are primarily built on the Transformer architecture, a deep learning framework that uses the “attention” mechanism. This architecture has become especially prominent since Google introduced BERT in 2018.

There are three main types of Transformer models: 1) Encoders -- These process input data (like text) and produce dense representations or embeddings; 2) Decoders -- These generate new tokens sequentially, predicting one token at a time to form coherent output; 3) Encoder-Decoder (Seq2Seq) -- This setup first uses an encoder to process the input sequence into a contextual representation, which a decoder then uses to generate an output sequence. Although Transformers come in various forms, most LLMs today are decoder-based, designed for text generation tasks and composed of billions of parameters. The core concept behind LLMs is straightforward yet powerful: they aim to predict the next token in a sequence, given the preceding tokens. A token is the fundamental unit of information an LLM operates on. While it’s similar to a “word”, LLMs typically use sub-word tokens for efficiency. Text generation remains the most common application of LLMs. Given an initial input or prompt, the model predicts one token at a time, continuing the sequence until it reaches a specified length or encounters an end-of-sequence (EOS) token. 
Considering their robust summarization and generation capabilties, in automotive, LLMs are adopted for various tasks within software development toolchain, including requirements analysis and summarization, executable code and test generation, analysis and optimization.
\subsection{Prompting Techniques}
Prompt engineering is crucial for guiding LLMs in generating accurate, relevant, and functional code. LLMs like GPT-4 are autoregressive models trained to predict the next token based on previous context. This means they generate statistically likely continuations, not reason symbolically like traditional programming tools \cite{brown2020language}. When they are asked to write code, they’re using statistical patterns from billions of examples of text and code to synthesize something that looks like a correct response. Without a well-designed prompt, the model might misinterpret the task, skip important constraints, hallucinate functionality, or generate incomplete or nonfunctional code \cite{liu2021pretrain}.

Prompt engineering involves crafting the input (prompt) so that the LLM \cite{liu2021pretrain}:
\begin{itemize}
  \item Understands context and goals (e.g., "This is code for a self-driving car system.").
  \item Uses appropriate reasoning steps (especially with CoT and ReAct).
  \item Maintains structure (e.g., “Write a function, then write tests.”).
\end{itemize}

Commonly used prompting techniques are:
\begin{itemize}
 \item Direct Prompting: Gives a simple instruction without reasoning steps. Strengths: Fast and simple. Weaknesses: Prone to logical errors and misses edge cases.

\item Chain of Thought (CoT): Prompts the model to explain intermediate reasoning before generating code. Strengths: Improves logic and transparency. Weaknesses: More verbose and slower.

\item ReAct (Reason + Act): Alternates between reasoning and actions, simulating agent-like behavior. Strengths: Suited for interactive, step-by-step tasks. Weaknesses: Complex to design and debug.

\item Retrieval-Augmented Generation (RAG): Injects external knowledge (e.g., regulations, documentation) into prompts dynamically. Strengths: Enables domain-grounded and accurate code. Weaknesses: Dependent on retrieval quality and adds infrastructure complexity.

\item Long-Duration Autonomy Prompting: Designed for extended sessions where context must persist over time. Strengths: Supports long, consistent development workflows. Weaknesses: Risk of context drift and higher system demands.
\end{itemize}

\subsection{Retrieval Augmented Generation (RAG)}

Typical RAG-enabled workflow consists of two main processes: 
1) indexing -- involves creation of a pipeline to ingest and index data from a textual source (such as PDF or Word document), typically performed apriori, offline. Furthermore, this phase can be broken down into the following steps: a) load - data is initially loaded using document loaders; b) split - large documents are broken into smaller chunks with text splitters, making them easier to search and fit within the model's context window; c) store - these chunks are stored and indexed, often using a vector store and an embeddings model, for efficient searching later.
2) retrieval and generation -- runtime process where the system takes a user query, retrieves relevant data from the index, and passes it to a model for generating a response. This phase consists of: a) retrieve -- relevant data chunks are fetched from storage using a Retriever in response to a user query; b) generate -- LLM generates an answer by using a prompt that combines the question and the retrieved data. \\
In the context of automotive software development, RAG plays a pivotal role in addressing the increasing complexity of regulatory compliance and RFQ (Request for Quotation) processing. As OEMs face mounting pressure to align technical specifications and manufacturing plans with frequently evolving standards—such as Euro 7 emissions regulations—manual tracking and analysis become error-prone and time-consuming. By combining RAG with AI agent networks, companies can automate the retrieval of relevant regulatory documents, extract design-related constraints, and assess compliance dynamically \cite{meng_using_2025} \cite{amazon_bedrock}. This approach not only reduces the risk of overlooking critical requirements but also enables faster and more accurate responses to RFQs, ultimately supporting more agile and regulation-aware development cycles in the automotive domain.
\subsection{Vision-Language Models (VLMs)}
Vision-Language models are multimodal systems aiming to learn from both images and text simultaneously, which gives them the ability to include information originating from visual representations within the generated response. This way, it is possible to perform variety of tasks—from visual question answering to image captioning. At their core, vision language models are generative models that process both visual and textual inputs to produce text-based outputs. These models are capable of handling various image types—including documents and web pages—and demonstrate strong zero-shot performance and generalization abilities.

When it comes to automotive, they support a broad range of use cases, such as extracting and analyzing requirements represented within illustrations, diagrams and graphs, which is of particular importance in automotive as well, considering that precious information is usually present in this form as well within relevant documents (vehicle documentation, standards). Additionally, some models are even capable of object recognition and understanding of spatial relationships within an image, allowing them to generate bounding boxes, segmentation masks, or answer questions about the location of objects, making them adoptable for complex autonomous driving-related scenarios \cite{zhou2024vision}. However, within the scope of this paper we focus on automotive software development use cases, while sensor data fusion and autonomous driving capabilities implementation relying on VLMs is outside the scope of this work.

%% file: sections/02b_methodology.tex
\begin{figure*}[h]
	\includegraphics{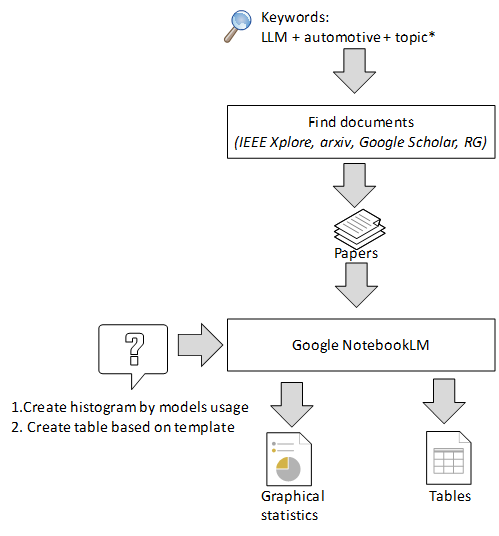}
	\caption{Underlying methodology for each of the topics covered used for this survey paper creation}
	\label{fig:method}
\end{figure*}

This sections describes the underlying methodology applied for creation of individual paper sections. Depiction of paper retrieval and processing workflow is given in Fig. \ref{fig:method}. In the first step, it is assumed that human manually looks up the paper sources, such as IEEE Xplore and ResearchGate, as well as indexing services like Google Scholar and preprint platform such arxiv. For that purpose, we used the following search pattern for the keywords: LLM+automotive+topic*, where topic is specific for the paper section, such as requirements, code generation, compliance, test generation, test scenarios. Considering that adoption of GenAI in area of automotive is still in not mature, due to lack of papers targeting this particualr domain, we also include some of more general research works leveraging GenAI for software engineering-related tasks (requirements handling, modeling, code generation) that can be potentially adopted in case of automotive as well. However, we assume that not all the papers are taken automatically from the search results, but are curated by human reader and filtered. Once we get the paper files in PDF format, we further process them by uploading them to Google's NotebookLM service \cite{notebooklm}. The main functionality of this service is to summarize group of documents and enable convenient prompting against the summarized information also relying on LLMs. This way, we can conveniently extract summarized reports, such as tables according to given template or histograms/usage statistics for distinct models within the given topic (like most used LLMs for automotive code generation and similar).

%% file: sections/03_workflow.tex
In this section, we present the generalized workflow of GenAI-empowered approach to automotive software development, based on both the existing literature and our 2-year work on CeCaS project \cite{mannheim_cecas_tum} \cite{lebioda2024singlesystem} \cite{petrovic2024synergy}. Fig. 1 depicts the generalized workflow. For each of the steps within the workflow, the label denoting corresponding GenAI methodology usually applied, as well as auxiliary techniques where needed (such as Model-Drive Engineering).
In the first step, the provided documents - either customer-specific requirements or regulations (such as standards) are processed in chunks relying in RAG, in order to enable efficient retrieval of relevant information which is leveraged for construction of underlying datasets for code and test generation. In this context, regulation documents such as UN157 are often used as basis for test scenario generation using RAG. However, considering that visual representations, such as diagrams and graphs capture valuable knowledge in automotive domain documentation (either requirements or standards and regulations), VLMs can be also adopted in order to extract the relevant information. Moreover, usually the next step towards any code generation is leveraging formal representation \cite{petrovic2024synergy} \cite{abdalla2024}) as intermediary product before code generation. For that purpose, LLMs are used to summarize the extracted requirements with respect to some formal template or metamodel. This way, it is more convenient to perform additional checks in design-time, such as compliance with standards. Once the requirements completeness, correctness and compliance checks are finished, code can be further generated using LLMs. Considering the nature of automotive domain, usually the simulation code is first generated, so the functionality can be evaluated before reaching the target platform. When it comes source code, LLMs can be leveraged to perform its analysis with respect to various aspects relevant for automotive, such as security compliance, source code-level safety and MISRA C standard compliance. However, usually it is considered that there is also human expert/reviewer in loop, whose role is to provide feedback in case that some inconsistencies or wrong results are generated. At the current state-of-art, considering the sensitivity of automotive domain itself, there are still steps within GenAI-based workflows that assume human experts presence and their feedback. However, it can be also noticed that novel approaches emerge in order to reduce the amount of human intervention needed by incorporating approaches that rely on multiple LLM agents, such as \cite{schamschurko2025recsip}. 
\begin{figure*}[h]
	\centering
	\includegraphics[width=\linewidth]{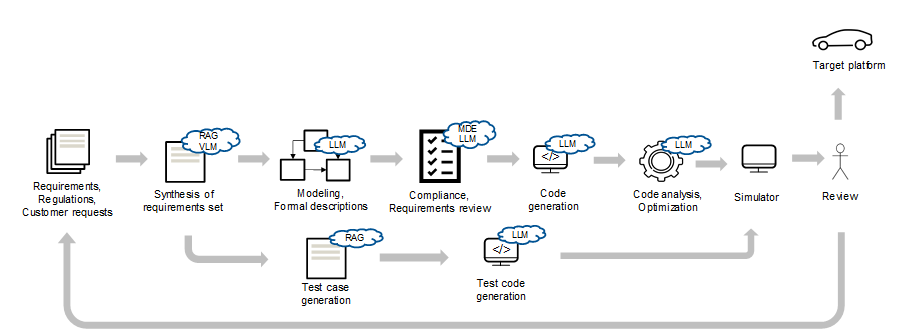}
	\caption{GenAI-driven automotive software development workflow}
	\label{fig:workflow}
\end{figure*}

%% file: sections/04_requirements.tex
Considering the fact that single vehicle consists hundreds thousands of requirements, it can be concluded that any manual process involving their consideration and analysis requires domain expertise, as well as enormous amount of time and effort. Therefore, handling such number of requirements using GenAI, especially LLMs and RAG is identified one of crucial aspects which leads to much faster development cycles. 

Some of the solutions focus on question answering based on given requirements documents \cite{zolfaghari2024rag}\cite{uygun2024}, as well as model-based software automotive system representations\cite{mazur2025querying}. On the other side, there are also approaches that aim to produce software artifacts which can be further leverage for compliance checking and code generation, such as model instances with respect to metamodel (such as Eclipse Modeling Framework's Ecore) or schema (such as JSON) \cite{petrovic2025meta} \cite{pan2025model} . Based on the existing works, it can be noticed that for this type of tasks, locally deployable, fine-tuned models are preferable (such as fine-tuned Llama 3), which was expected considering the non-disclosure policies of the underlying organizations. Table \ref{tab:rel3} summarizes relevant works, use cases and models.

\begin{table*}[htbp]
\caption{LLM and RAG-based requirements handling in automotive}
\begin{center}
\scriptsize
\begin{tabularx}{\textwidth}{|c|X|X|X|X|}
\hline
\textbf{Reference} & \textbf{Use case} & \textbf{Input} & \textbf{Output} & \textbf{Model} \\
\hline
Uygun et al. (2024) \cite{uygun2024} & Question answering and analysis & Requirements and regulatory documents & Textual answers & \begin{itemize} \item Nous-Hermes-13B-GPTQ \item WizardLM-7B-Uncensored-GPTQ \item Wizard-Vicuna-13B-Uncensored-GPTQ \item guanaco-7B-GPTQ \item orca\_mini\_v2\_13b-GPTQ \end{itemize} \\
\hline
Petrovic et al. (2025-1) \cite{petrovic2025meta} & Metamodel construction & OEM requirements document as freeform text & Ecore EMF metamodel & \begin{itemize} \item GPT-4o \item deepseek-ai/deepseek-llm-7b-chat \end{itemize} \\
\hline
Pan et al. (2025) \cite{pan2025model} & Automtoive model instance creation & User-specified requirements as freeform text & XMI model instance & \begin{itemize} \item GPT-4o \item o1-preview \item Llama 3.1 (8B and 70B) \end{itemize} \\
\hline
Pan et al. (2024) \cite{pan2024ocl} & Requirements completeness and correctness check & Regulations and reference architecture requirements as freeform texts & Object Constraint Language (OCL) rules & \begin{itemize} \item GPT-4 Turbo \item Gemini 1.5 Pro \item Llama 3 8B (with and without fine-tuning) \end{itemize}\\
\hline
Zolafghari et al. (2024), Li et al. (2025) \cite{zolfaghari2024rag} \cite{li2025optimizing}& Compliance & Regulations and standardization documents & Freeform textual answer & \begin{itemize} \item GPT-4o \item LLAMA3 \item Mistral \item Mixtral \end{itemize}\\
\hline
Mazur et al. (2025) \cite{mazur2025querying} & Querying large automotive model instances & Model instances & Freeform textual answer & \begin{itemize} \item GPT-4o mini \item GPT-4.1 mini \item o4-mini \item Gemini 2.5 Flash Preview \end{itemize}\\
\hline

Obstbaum et al. (2024) \cite{obstbaum2024specbook} & Requirements formalization and quality control & Raw requirements & Formalized and refined requirements; test cases & \begin{itemize} \item GPT-4 \item Llama3 \end{itemize}\\
\hline

\end{tabularx}
\label{tab:rel3}
\end{center}
\end{table*}

%% file: sections/04a_compliance.tex
This section explores recent developments in applying Large Language Models (LLMs) and Retrieval-Augmented Generation (RAG) for regulation-compliant scenario generation in the automotive domain. While there is a growing body of research on using LLMs for general compliance verification, this survey concentrates specifically on the use of LLMs and RAG techniques for generating test scenarios that support scenario-based testing of Autonomous Driving Systems (ADS). We review and evaluate the existing literature from the perspective of the underlying source documents—such as accident reports, technical regulations, and other sources. For each approach, we briefly summarize the proposed solution, highlight how regulatory texts are processed or interpreted, and analyze how effectively the models align the generated scenarios with the intent and structure of the referenced documents.
To structure our analysis, we categorize the reviewed works based on the transformation between different scenario abstraction levels. We distinguish three types of scenarios:
\begin{itemize}
    \item Functional scenarios, which describe a situation at a high semantic level by outlining the involved entities and their interactions.
    \item Logical scenarios, which refine the functional descriptions by introducing parameterized state spaces with defined value ranges.
    \item Concrete scenarios, which instantiate logical scenarios by assigning specific values to each parameter.
\end{itemize}
By mapping the reviewed studies to these abstraction levels and analyzing how they convert between them, we aim to provide a clearer understanding of how LLMs and RAG systems contribute to the end-to-end process of regulation-compliant scenario generation.
Chat2Scenario \cite{zhao_chat2scenario_2024} proposes a pipeline that extracts concrete driving scenarios from naturalistic datasets using GPT-4, focusing on criticality thresholds and translating the outputs into simulation-ready formats like OpenSCENARIO. TARGET \cite{deng_target_2023} leverages GPT-4 to parse traffic rules written in natural language and converts them into a formal domain-specific language (DSL) for structured test scenario generation. LEADE \cite{tian_lmm-enhanced_2025} employs vision-language models and traffic videos to reconstruct safety-critical scenarios through behavior comparison between human drivers and autonomous systems. Lastly, LeGEND \cite{tang_legend_2024} introduces a top-down approach for transforming textual accident reports into structured functional, logical, and concrete scenarios using a two-phase LLM-based transformation. \ref{fig:method} is a synthesized general pipeline that abstracts the key steps common across these solutions.

\begingroup
\renewcommand{\arraystretch}{1.4}

\begin{table*}[t]
\centering
\small
\begin{tabularx}{\textwidth}{lXXXX}
\toprule
\textbf{Feature} &
\textbf{Chat2Scenario \cite{zhao_chat2scenario_2024}} &
\textbf{TARGET \cite{deng_target_2023}} &
\textbf{LEADE \cite{tian_lmm-enhanced_2025}} &
\textbf{LeGEND \cite{tang_legend_2024}} \\
\midrule
\textbf{Source of Input} &
Naturalistic driving datasets &
Traffic rules (natural language) &
Real-world traffic videos &
Accident reports (natural language) \\
\midrule
\textbf{Scenario Understanding} &
LLM extracts scenario elements and converts them into structured representations. &
LLM parses rule-based descriptions and maps them to structured DSL syntax. &
LMM interprets vehicle behaviors from videos using optical flow analysis. &
LLM1 extracts accident events into Interactive Pattern Sequences (IPS). \\
\midrule
\textbf{Scenario Generation Method} &
Get scenarios from datasets, and converts to OpenSCENARIO \& CarMaker formats. &
Rule-based template filling for scenario definitions in DSL. &
Multi-modal few-shot CoT prompting to generate scenarios. &
Dual-LLM-powered transformation. \\
\midrule
\textbf{Final Output} &
Simulation-ready OpenSCENARIO \& CarMaker scenarios for ADS testing. &
Logical scenarios with parameterized ranges, converted into executable test cases. &
Concrete safety-critical scenarios for Baidu Apollo in SVL simulation. &
Concrete crash scenarios tested in Baidu Apollo ADS. \\
\bottomrule
\end{tabularx}
\caption{Comparison of LLM- and RAG-based Scenario Generation Approaches for Automotive Regulation Compliance}
\label{tab:llm_scenario_comparison}
\end{table*}

\endgroup

\begin{figure*}[h]
	\centering
	\includegraphics[width=\linewidth]{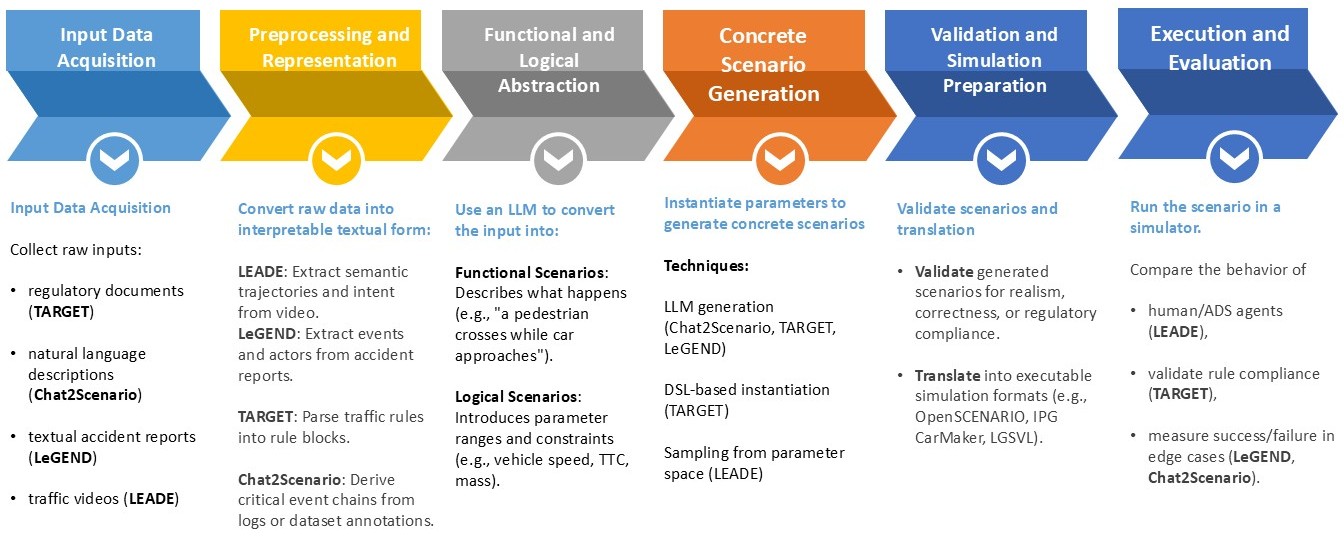}
	\caption{Summary of Scenario Generation solutions}
	\label{fig:scenario_generation}
\end{figure*}

%% file: sections/06_code_generation.tex
The summary of relevant existing works leveraging LLMs for code generation in area of automotive is given in Table \ref{tab:rel1}. In general, it can be noticed that there are two major distinctive approaches: 1) direct - aims to robustly generate code starting from textual descriptions 2) indirect - introduces additional intermediate representations, such as model based representations in order to introduce the ability to verify model-level correctness in design-time.
In \cite{patil2024}, two variants of commercial OpenAI's GPT (3.5 and 4-turbo) were used for purpose of MISRA-compliant C code generation, starting from various aspects of specification. The specification in this work can be provided as either freeform text high-level or low-level descriptions, while formal ANSI/ISO C Specification Language (ACSL) is supported as well. 
In \cite{Liu2024}, the authors experimented with GPT 4 for generating safety critial software as C code. Leveraging the power of LLM, they also proposed an adapted software development process and lifecycle V-model.
On the other side \cite{nouri2025} aims generation of Python code for simple esmini simulator, where various commercial models were considered. Additionally, in \cite{abdalla2024}, the authors have adopted fine-tuned variants of smaller LLM instances (such as Code-Llama 7B) and compared them to commercial GPT-based solutions, while the target C code was generated in several steps, starting from textual software requirements and leveraging MATLAB Simulink as intermediate model representation. On the other side, in \cite{pan2025automating}, \cite{lebioda2025requirements} make use of GPT4-based models in order to generate target code for CARLA simulator, covering the aspects of vehicle configuration (sensors and actuators), as well as its behavior.

\begin{table*}[htbp]
\caption{LLM-based code generation in automotive}
\begin{center}
\scriptsize
\begin{tabular}{|c|p{3cm}|p{3cm}|p{2cm}|p{2cm}|}
\hline
\textbf{Reference} & \textbf{Use case} & \textbf{Input} & \textbf{Target} & \textbf{Model} \\
\hline
spec2code \cite{patil2024} & 1) oil level warning 2) brake light activation 3) power steering backup & Freeform text, ACSL & ANSI C & GPT-3.5 and GPT-4-turbo \\
\hline
Nouri et al. (2025)\cite{nouri2025} & Adaptive Cruise Control (ACC) and Unsupervised Collision Avoidance by Evasive Manoeuvre (CAEM). & Function and scenario description - OpenSCENARIO and OpenDRIVE format  & Python for esmini simulator & Codellama:34b, DeepSeek (r1:32b and Coder:33b), CodeGemma:7b, Mistral:7b, GPT-4 \\
\hline
Abdalla et al. (2024) \cite{abdalla2024} & Simple automotive function development &  Software requirements & Matlab model,  code documentation & Mistral 7B, Code-Llama 7B (fine tuned), GPT-3.5-turbo, GPT-4 \\
\hline
Liu et al. (2024) \cite{Liu2024} & Cruise Control & Software Requirements & C & GPT-4 \\
\hline
Pan et al. (2025-2) \cite{pan2025automating} & Emergency braking & Vehicle behavior description & Python - CARLA simulator code, ROS2 as middleware & GPT-4o \\
\hline
Lebioda et al. (2025), Petrovic et. al (2024-2) \cite{petrovic2024llm} & Emergency braking & Vehicle configuration - sensor and actuator configuration & Python - CARLA simulator code & GPT-4o \\
\hline
\end{tabular}
\label{tab:rel1}
\end{center}
\end{table*}

%% file: sections/07_hallucination.tex
First, we have to define what a hallucination is and which types we would want to tackle.
Following~\cite{Ji_2023}, hallucinations are generated content that is nonsensical or unfaithful to the provided source content.
As we focus on the application of LLMs and VLMs for the automotive software development in this paper, we differentiate between the data format of the hallucinations. 
Therefore, we focus on methods for reducing or recognizing textual and visual hallucinations.

\subsection{Textual hallucinations}
One common reason for hallucination in reasoning and self-reflection is the Degeneration-of-Thought (DoT) problem, which was proposed and defined by Liang et al.~\cite{liang2024encouragingdivergentthinkinglarge}.
It describes the problem of LLM-based agents having a high confidence in their answer although it was initially incorrect.

The approaches for these hallucinations differ in the number of used language models, in the correction approach and in the scoring of the output to decide which outputs could be hallucinations.
A summary of these can be found in~\cref{tab:summaryTextualHallucinations}.

\begin{table}
    \caption{Summary of the methods tackling textual hallucinations}
    \label{tab:summaryTextualHallucinations}
    \begin{tabularx}{\linewidth}{X|c|c}
        \textbf{Approach} & \textbf{Multiagent} & \textbf{LLM as a Judge} \\
        \hline\hline
        Multiagent debate~\cite{liang2024encouragingdivergentthinkinglarge,chen-etal-2024-reconcile,wang-etal-2024-rethinking-bounds} & yes & yes \\
        \hline
        Self-Consistency~\cite{wang_self-consistency_2023,wan_dynamic_2024,liang_internal_2024,li2024escape,aggarwal2023letssamplestepstep} & no & no \\
        \hline
        RECSIP~\cite{schamschurko2025recsip} & yes & no \\
        \hline
        Uncertainty Evaluation~\cite{manakul_selfcheckgpt_2023,zheng2024evaluating,wang_clue_2024} & no(yes) & yes \\
    \end{tabularx}
\end{table}

\subsubsection{Multiagent debate}
The focus of the multiagent debate (MAD) framework~\cite{liang2024encouragingdivergentthinkinglarge} is on the discourse between disparate LLM agents, with the understanding that the output of one agent becomes the input of another.
This approach involves the utilisation of multiple LLMs as debaters and a single LLM as the judge of the debate.
The debaters have been assigned defined roles, categorised as either affirmative or negative debaters.
The objective of the framework is to enable the debaters to articulate their respective viewpoints, supported by a rationale, and subsequently, the judge LLM will determine the winning side of the debate.

Chen et al. introduce ReConcile~\cite{chen-etal-2024-reconcile}, a multiagent group discussion with confidence estimation.
In the course of their argument, each debater defends their position and presents their argument in support of it.
They then offer their own estimation of their confidence.
As the participants continue to engage with each other's arguments, the confidence score undergoes fluctuations during the course of the discussion.
The result of the discussion is reached when the confidence scores attain a predetermined threshold.

Wang et al.~\cite{wang-etal-2024-rethinking-bounds} re-evaluate the claim that multi-agent discussions improve the reasoning abilities of Large Language Models (LLMs).
They introduce a new multi-agent discussion framework called Conquer-and-Merge Discussion (CMD), inspired by human group discussion processes.
CMD aims to provide a more comprehensive comparison between single-agent and discussion frameworks.
The CMD framework consists of three stages: group discussion, voting and a final decision stage (used in case of a tie).
The LLM agents are divided into groups, share information within their group (including explanations) and only viewpoints from agents in other groups during the discussion rounds.
In the voting stage, each agent's vote is treated equally, with a majority decision determining the result.
In their performance evaluations the authors indicate that a single LLM with a strong prompt can perform comparably to multiagent debates (MAD and CMD).

\subsubsection{Self-Consistency}
Self-Consistency~\cite{wang_self-consistency_2023,wan_dynamic_2024,liang_internal_2024,li2024escape,aggarwal2023letssamplestepstep} is a decoding strategy designed to improve the reasoning capabilities of large language models.
Rather than relying on a single greedy approach, this method enhances the existing chain-of-thought prompting by sampling multiple diverse reasoning paths.
Self-Consistency significantly boosts performance across various arithmetic, commonsense and symbolic reasoning tasks by aggregating the most consistent answer from these sampled paths.
It even outperforms prior approaches that require additional training or human annotations.
Research demonstrates the robustness of Self-Consistency across different model scales and sampling strategies, even with imperfect prompts, suggesting its utility in improving model accuracy and providing uncertainty estimates.

\subsubsection{REpeated Clustering of Scores Improving the Precision (RECSIP)}
Due to the stochastic nature of LLMs, their responses can be unreliable, posing risks in high-stakes environments.
RECSIP~\cite{schamschurko2025recsip} addresses this by querying multiple LLMs in parallel, then scoring and clustering their responses to identify a more trustworthy answer.
The framework employs a callback mechanism where, if disagreement exists, LLMs are prompted with a multiple-choice question derived from the clustered responses to reach a consensus. 

\subsubsection{Uncertainty Evaluation}
The field of uncertainty evaluations varies from sequence level uncertainty~\cite{manakul_selfcheckgpt_2023,zheng2024evaluating} to concept level uncertainty~\cite{wang_clue_2024}.
All uncertainty evaluations are integrated to quantify the stochastic process of LLMs.
The majority of works in this field employ a second LLM for this purpose or allow the LLM to quantify its own response.
Each approach assigns a numerical value to the LLM output, which is then interpreted by the user or a framework to determine subsequent action.

\subsection{Visual hallucinations}

%% file: sections/08_code_optimization.tex
Code generated by Large Language Models (LLMs), often requires refinement beyond initial syntactic and functional correctness. This section focuses on analyzing such generated code and applying targeted optimization strategies to improve performance, maintainability, and compliance with safety standards. After initial generation and validation, code quality can be significantly improved by enforcing best practices and design patterns, such as adherence to MISRA C/C++ guidelines or standards like ISO 26262 \cite{ISO26262}. These checks go beyond correctness and focus on maintainability, readability, and safety-critical reliability. Semantic optimization includes simplifying logic structures, reducing nested conditions, and improving modularization. Traditional tools such as cppcheck \cite{cppcheck2024} identify redundant code paths, unused variables, and type-related inefficiencies. Moreover, incorporating domain-specific patterns, such as efficient fixed-point arithmetic in embedded systems, can yield performance benefits under constrained hardware.

While functional testing verifies correctness, runtime profiling identifies performance bottlenecks. In automotive, generated code is usually first executed in a simulation environment where real-time performance metrics—such as CPU/GPU load, memory footprint, and execution latency—are captured. Profiling tools assist in identifying bottlenecks, allowing targeted refactoring or prompting strategies to produce more efficient implementations in the next iteration. For instance, in the Adaptive Cruise Control (ACC) use case, profiling revealed that certain distance calculations could be simplified, reducing unnecessary computations per simulation frame. This insight is integrated into the next iteration of LLM prompting, thereby improving both generation quality and runtime efficiency.

In \cite{zhang2024}, to optimize generated code, feedback from static analysis and runtime profiling are taken back into the prompting loop. This iterative refinement process uses error messages, performance metrics, and optimization suggestions as contextual input for the LLM. Chain-of-Thought prompting enables the model to reason through the implications of specific changes, e.g., reducing branching depth or eliminating repeated calculations. 

Additionally, when the generated code is non-optimal but correct, role-based prompting can re-frame the LLM as an “automotive optimization expert” to suggest domain-specific improvements \cite{kong2024}. This dual use of the LLM—as both a code generator and refactoring assistant—enables systematic improvement across multiple software development dimensions.

Summary of notable works on code optimization in automotive with underlying models applied for those tasks are given in Table \ref{tab:rel4}.

\begin{table*}[htbp]
\caption{LLM-based code optimization}
\begin{center}
\scriptsize
\begin{tabular}{|c|p{4cm}|p{4cm}|p{3cm}|}
\hline
\textbf{Reference} & \textbf{Aspect} & \textbf{Target/Scenario} & \textbf{Models} \\
\hline
LangProp (2024) \cite{ishida2024langprop} & Iterative, metrics-based and data-driven code optimizations & CARLA code in Python for autonomous driving capabilities &  GPT 3.5 Turbo 16k \\
\hline
Sevenhuijsen et al. (2025) \cite{sevenhuijsen2025} & Automating Generation of Formally Verified C Code with Large Language Models & Software requirements to formally verified C Code & Llama-3.1-70B, GPT-3.5-turbo, and GPT4o\\ 
\hline
Purschke et al. (2025) \cite{purschke2025speedgen} & Performance bottleneck identification & Python code & Llama 3.1-8B Instruct\\ 
\hline
Kirchner et al. (2025) \cite{kirchner2025generating} & Safety analysis and optmization & C++ code for assisted driving & Qwen2.5-Coder-7B-I, Llama-3-8B-Instruct , DeepSeek-Coder-V2 LI , DeepSeek-Coder 33B I , CodeStral-22B , GPT-4o (240513) \\ 
\hline

\end{tabular}
\label{tab:rel4}
\end{center}
\end{table*}

%% file: sections/09_vlm_reqs.tex
In general, automotive software development relies on hundred thousands of requirements and comprehensive documentation, which usually includes not only textual information, but visual information as well. Various diagrams, graphs and schematics are adopted in order to encapsulate various aspects of vehicular system, from purely software-related elements, to its high-level behavior. Despite that modern model-driven tools such as AVL Cameo \cite{avl_cameo5} and other SysML-based tools \cite{sysml2025} enable storing the created diagrams in structured textual representations (with respect to metamodel or schema), there are several possible situations when still the extraction directly from visual format is needed \cite{petrovic2025vlm}. One of the cases is backwards compatibility, updating and maintenance of older products, so original diagram files might be lost or the tools themselves were purely drawing-oriented (rather than modeling). On the other side, in case of regulations and standards, some of the information and values about test scenarios (such as UN152 and UN157 regulations for emergency braking and lane keeping systems) are represented within freeform illustrations, so we cannot rely on structured textual representations of the images. 

Therefore, adoption of multimodal summarization, more precisely VLMs is one of key enablers in requirements dataset synthesis step in GenAI-driven automotive software development \cite{petrovic2025vlm}. In Table \ref{tab:rel2}, the summary of the relevant works on this topic is given. It can be noticed that most of the existing solutions make use of UML notation diagrams (usually class diagrams \cite{petrovic2025vlm}, \cite{munde2025}) to extract information which is possibly further leveraged - for system updates, code or test generation. When it comes to underlying models, it can be noticed that both commercial and locally deployable, smaller instances are also used. For this kind of tasks related to diagram handling in automotive, in most cases the solutions rely on pre-trained models, as they seem to exhibit acceptable performance, even for smaller number of parameters. On the other side, when it comes to understanding the order and sequence aspects within flowchart-alike and scientific diagrams, in most cases additional refinements are needs, such as intermediary reasoning steps \cite{ye2024beyond} or even synergy with traditional computer vision methods \cite{omasa2025, li2025chainofregion} in order to ensure acceptable result accuracy.
\begin{table*}[htbp]
\caption{VLM-based diagram handling in automotive}
\begin{center}
\scriptsize
\begin{tabular}{|c|p{4cm}|p{4cm}|p{3cm}|p{3cm}|}
\hline
\textbf{Reference} & \textbf{Use case} & \textbf{Input} & \textbf{Output} & \textbf{Model} \\
\hline
Petrovic et al. (2025-2) \cite{petrovic2025vlm} & Software updates & UML class diagrams as images & Automotive software update commands & Grok-2, Pixtral-12B, Claude-3.5, Gemini-2.0, Gemini-1.5, GPT-4o, GPT-o1, GPT-4, GPT-4o-mini, InternVL2-8B-MPO, Qwen2 VL 7B \\
\hline
Munde (2025) \cite{munde2025} & Parameter extraction & UML class diagrams as images & Question answering about flowchart elements (paremeters, order) & Qwen2VL, LLAVA 1.6, Kosmos-2, GPT-4o-mini \\
\hline
Ye et al. (2024) \cite{ye2024beyond} & Question answering & Flowchart diagrams from datasets: TEXTFLOW, FlowVQA, FlowLea & Question answers about UML parameters & GPT-4o (VQA), Claude3.5 Sonnet, Qwen2.5-32B, Llama3.1-70B, Qwen2.5-14B \\
\hline
Omasa et al. (2025) \cite{omasa2025} &  Diagram element connection analysis & Flowchart and graph-alike diagrams  & Arrow directions & DAMO-YOLO \\
\hline
Li et al. (2025) \cite{li2025chainofregion} &  Diagram analysis & Scientific diagrams from various areas: biology, chemistry, geography, mathematics and phyiscs & Question answering & GPT-4o, GPT4-mini, GPT4-Turbo \\
\hline
\end{tabular}
\label{tab:rel2}
\end{center}
\end{table*}

%% file: sections/10_prompt.tex

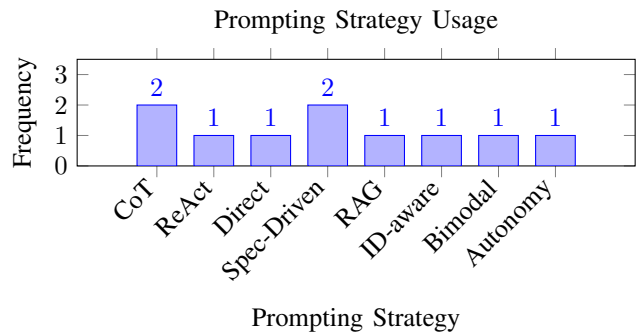
\begin{figure}[h!]
\centering
\begin{tikzpicture}
\begin{axis}[
    ybar,
    symbolic x coords={CoT, ReAct, Direct, Spec-Driven, RAG, ID-aware, Bimodal, Autonomy},
    xtick=data,
    ylabel={Frequency},
    xlabel={Prompting Strategy},
    ymin=0, ymax=3.5,
    bar width=15pt,
    width=9cm,
    height=3cm,
    nodes near coords,
    enlarge x limits=0.2,
    xticklabel style={rotate=45, anchor=east},
    title={Prompting Strategy Usage}
]
\addplot coordinates {
    (CoT,2)
    (ReAct,1)
    (Direct,1)
    (Spec-Driven,2)
    (RAG,1)
    (ID-aware,1)
    (Bimodal,1)
    (Autonomy,1)
};
\end{axis}
\end{tikzpicture}
\caption{Histogram of Prompting Strategy Usage Across Reviewed Works}
\label{fig:prompting-strategies}
\end{figure}

Prompting strategies shape how LLMs interpret and decompose the task. Techniques like Chain of Thought (CoT), Specification-Driven, ReAct, and Retrieval-Augmented Generation (RAG) vary in how they elicit reasoning. The bar chart in Fig.~\ref{fig:prompting-strategies} reflects the frequency of each strategy observed in the reviewed studies. Notably, Specification-Driven prompting and CoT appear frequently in automotive-focused tasks, due to the structured and logic-intensive nature of the domain. Additionally, two more prompting techqniues are also identified within the automotive-related works: identifier-aware and specification-driven prompting.

Identifier‑Aware prompting (or pre‑training) teaches a model to explicitly recognize and recover code identifiers during generation. CodeT5 introduced identifier‑tagging and masked‑identifier prediction during pre‑training, allowing superior handling of variable/function names in code generation tasks~\cite{wang2021codet5, wang2021codet5}. This labeling ability improves semantic consistency across generated code.  

Strengths:
\begin{itemize}
    \item Generates semantically coherent code with preserved variable semantics.
    \item Reduces tokenization ambiguities.
\end{itemize}

Weaknesses:
\begin{itemize}
    \item Benefits depend on pre-trained capabilities.
    \item Less useful in unstructured prompts.
\end{itemize}

Specification‑Driven prompting consists of embedding explicit specifications (formal, semi-formal, or natural-language requirements) into the prompt to guide code generation. In \cite{petrovic2024llm} \cite{petrovic2024synergy} \cite{lebioda2025requirements}, we previously used Ecore and JSON model specifications to generate verified code for emergency braking systems in CARLA simulations~\cite{petrovic2024synergy}. Similarly, Patil et al.\ (2024) generated embedded cruise-control C code directly from formal specs~\cite{patil2024spec2code}.  

Strengths:
\begin{itemize}
    \item Produces verifiable, domain-conforming code.
    \item Aligns well with regulated environments.
\end{itemize}

Weaknesses:
\begin{itemize}
    \item Requires detailed spec writing.
    \item Increases engineering overhead.
\end{itemize}

\begin{table*}[h]
\centering
\caption{Summary of Prompting Strategies in Automotive Use Cases}
\label{tab:llm_usecases_1}
\begin{tabular}{|p{2cm}|c|p{2cm}|p{2cm}|p{1.4cm}|p{1.2cm}|p{3cm}|p{3cm}|}
\Xhline{3\arrayrulewidth}
\textbf{Reference} & \textbf{Auto} & \textbf{Aspect} & \textbf{Use Case} & \textbf{Prompting} & \textbf{Model(s)} & \textbf{Outcome} & \textbf{Target Output} \\ \Xhline{3\arrayrulewidth}
Petrovic et al. (2024)~\cite{petrovic2024synergy} & Yes & Vehicle Safety Systems, Centralized Vehicle Systems & Emergency Brake Scenario & Specification-Driven & GPT-based LLMs & Generated code evaluated in CARLA simulator & Ecore model representation + Python/CARLA code \\
\hline
Shi et al. (2024)\cite{shi2024aegis} & Yes & Functional Safety Engineering & HARA process generation & RAG & Aegis-Max & Automated functional safety artifact generation & Hazard logs + test artifacts from ISO 26262 context \\
\hline
Patil et al. (2024)~\cite{patil2024spec2code} & Yes & Embedded Software and Cruise Control & Cruise Control logic synthesis & Specification-Driven & GPT-based LLMs & C code generated from natural and formal specs & Embedded C code auto-generated \\
\hline
Nouri et al. (2025)~\cite{nouri2025simulation} & Yes & Autonomous Driving Software & Adaptive Cruise Control (ACC) & ReAct & CodeLlama, DeepSeek, GPT-4 & Code validated in virtual test bench simulation & Python controller classes tested via simulation \\
\hline
Antero et al. (2024)~\cite{antero2024harnessing} & Yes & Robotics / FSMs & Control Logic for Autonomous Robot & Direct Prompting & GPT-based LLMs & Verified FSM transitions generated with LLMs & FSM code and logic graphs for simulation testing \\
\hline
Wang et al. (2021)~\cite{wang2021codet5} & No & Code Understanding & Multi-domain code synthesis & Identifier-Aware Prompting & CodeT5 & High-quality completions across domains & Python, Java, C++ snippets for benchmarks \\
\hline
Feng et al. (2020)~\cite{feng2020codebert} & No & Code Retrieval & Natural Language to Code & Bimodal Pre-Training & CodeBERT & Strong performance in code search & Ranked code results from natural language \\
\hline
Anthropic (2025)~\cite{anthropic2025claude} & No & General Code Generation & Long-duration coding tasks & Long-Duration Autonomy & Claude Opus 4 & Continuous multi-hour programming sessions & Working backend/frontend code via one prompt \\
\hline
Wei et al. (2022)~\cite{wei2022chain} & No & Reasoning Capability & Multi-step reasoning in coding & Chain of Thought & PaLM, GPT-3.5 & Showed improved logical structure in code generation & Problem decomposition into code blocks \\
\hline
Kojima et al. (2022)~\cite{kojima2022zeroshot} & Yes & Perception Reasoning & Lane detection logic for AD & Chain of Thought & GPT-3.5, GPT-4 & Multi-step perception pipelines synthesized via CoT & Modular sensor fusion code for AV systems \\ \\ \hline
\end{tabular}
\end{table*}

%% file: sections/10_survey.tex
In this section, we introduce survey involving partners from our ongoing CeCaS \cite{mannheim_cecas_tum} project related to future-proof automotive architecture development and implementation. The survey was created using Google Form. The goal of the questionnaire is to identify which are most common aspects that are already successfully automated, together with the underlying type of models (locally deployed or commercial services), as well as the overall user impression of the effectiveness of such tools. 

In what follows, the structure of the questionnaire is presented. 

\emph{Question 1: Is your organization using any GenAI-based tools?} 

\emph{Question 2: Does your organization rely on GenAI for requirements handling and automotive compliance?}\

\emph{Question 3: Does your organization rely on GenAI for code generation and/or augmentation?}

\emph{Question 4: Does your organization rely on commercial services and products (such as ChatGPT) or uses tailored models? }

\emph{Question 5: You work more efficiently relying on GenAI tools?}

\emph{Question 6: Type of organization?}

\begin{figure*}[h]
	\centering
	\includegraphics[width=\linewidth]{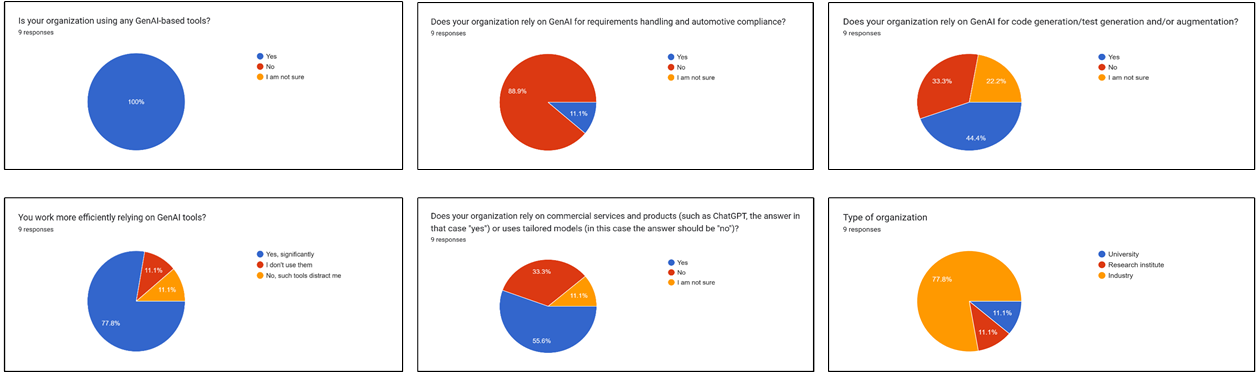}
	\caption{Usage of GenAI in automotive software development}
	\label{fig:survey}
\end{figure*}

For question 5, the possible answers are: a) Yes, significantly b) Yes, slightly c) I don't use them d) No, such tools distract me. Finally for question 6, one of the following answers had to be selected: a) University b) Research institute c) Industry. Summary of result outcomes is given in \ref{fig:survey}. The results included responses by 9 persons overall, 1 from university, 1 from research institute and 7 from industry.

Based on survey results, it can be concluded that all of the participants do rely on GenAI-based tools for software development. Major part of them is using such tools for some kind of code generation, while for requirements handling it was adopted only in one case. Majority agrees that such tools are useful and make our their work more productive. Most of the adopted solutions seems to be still based on commercial services.

Considering those results, it can be noticed that there is a gap when it comes to requirements handling using GenAI. This gap can be explained by the fact that most of the adopted solutions rely on commercial services which are not suitable for taking requirements as input, considering the fact that requirements as important asset of OEMs could hypothetically leave the bounderies of organizations this way. Therefore, for tasks related to direct handling of user stories and requirements, fine-tuned LLM-based solutions are the key enabler for increase of adoption.

%% file: sections/11_discussion.tex
\begin{figure}[h!]
\centering
\scriptsize 
\begin{tikzpicture}
\begin{axis}[
    ybar,
    symbolic x coords={GPT-4/GPT-4o, LLaMA, GPT-3.5, Mistral, DeepSeek, CodeLlama, Gemini, Mixtral, Others},
    xtick=data,
    ylabel={Frequency},
    xlabel={LLM Model Family},
    ymin=0, ymax=10,
    bar width=10pt,
    width=8cm,
    height=4cm,
    nodes near coords,
    nodes near coords align={vertical},
    enlarge x limits=0.1,
    xticklabel style={rotate=45, anchor=east},
    title={LLM Usage Across Reviewed Works}
]
\addplot coordinates {
    (GPT-4/GPT-4o,10)
    (LLaMA,8)
    (GPT-3.5,5)
    (Mistral,3)
    (DeepSeek,2)
    (CodeLlama,2)
    (Gemini,2)
    (Mixtral,1)
    (Others,6)
};
\end{axis}
\end{tikzpicture}
\caption{Histogram of LLM Usage Across Reviewed Works}
\label{fig:llm-usage}
\end{figure}
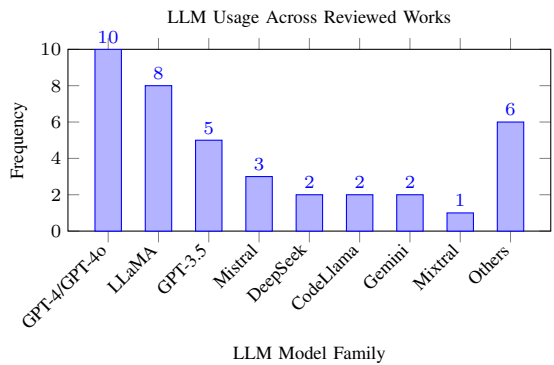

In Fig. \ref{fig:llm-usage}, a summary of LLM usage for dominant models across various automotive-related use cases is given. According to the results, it can be noticed that solutions based on OpenAI's commercial GPT model are still dominant, especially the latest variants. Such finding was expected, considering that GPT-based family of models are among the largest and most powerful ones, while their usage costs and are not locally deployable. However, considering those facts, their usage is more dominant in area of code generation, while for tasks directly handling requirements, fine-tuned and optimized variants of locally deployable models are used, especially Llama3 which is quite dominant among them. Such finding for this kind of use cases can be explained by the fact that most companies prefer not to expose requirements and user stories to third parties (such as external LLM services in cloud), as they are usually protected by NDAs and other similar contracts. However, adoption of locally deployable LLMs requires either time-consuming fine-tuning or additional strategies in order to reach acceptable quality of the results for domain such as automotive.

%% file: sections/12_conclusion.tex
In this paper, we provide state-of-art review of Generative Artificial Intelligence (GenAI) adoption within various steps of automotive software development toolchain - from requirements to executable code generation and optimization. The survey is based on around 60 publicly available research works. With respect to the current findings, it seems that automotive industry is beginning the adoption of GenAI-based tools, especially when it comes to code generation scenarios, where commercial solutions (such as GPT-4) are still dominant. In general, it seems that automotive industry has positive perception of GenAI adoption. Based on their experiences, users believe that such technology increases productivity, while reducing costs and complexity. Regarding the frequently used prompting techniques in automotive - CoT is dominant for behavior aspects-related code generation, while specification-driven methods are often preferred for simulation aspects, test scenarios and vehicle configuration scripts (sensors, actuators).

However, there is a research gap when it comes to requirements handling-related use cases of LLMs in automotive. Presence of such tools in automotive industry is lower, considering the aspects of intellectual property protection, where locally deployable LLMs are preferred. On the other side, adoption of such models requires huge datasets, powerful hardware for training, while the process itself is time consuming, making the adoption of such solutions much more challenging in practice. However, due to data protection policies and constraints affecting data sharing, the main bottleneck is lack of publicly available specialized automotive datasets which can be used for fine tuning.
In future, it is expected to see more publications on these topics, including fine-tuning strategies for LLMs tacking requirements handling. Our plan is to also  cover the aspects of hardware abstraction-related works aiming automotive systems, including interface and signal vehicle mapping with respect to given catalog using LLM s(such as Vehicle Signal Specification - VSS). Finally, we will also include complementary agentic AI-based approaches targeting scenarios, such as in-vehcile and infrastructure-related maintenance.

%% file: sections/13_ack.tex
This research was funded by the Federal Ministry of Research, Technology and Space of Germany as part of the CeCaS project, FKZ: 16ME0800K.